\documentclass[12pt,a4]{article}

\usepackage{tabularx}
\usepackage{graphicx}
\large
\title{Dating of two Paleolithic human fossil bones from Romania by accelerator mass spectrometry}
\author{Agata Olariu, G\"{o}ran Skog$^{\diamond}$, Ragnar Hellborg$^{\bullet}$, 
Kristina Stenstr\"om$^{\bullet}$,\\
Mikko Faarinen$^{\bullet}$,
Per Persson$^{\bullet}$,\\
Emilian Alexandrescu$^{\circ}$\\
{\em $^{}$Institute for Physics and Nuclear Engineering,}\\
{\em PO Box MG-6, 76900 Magurele, Bucharest, Romania}\\
{\em $^{\diamond}$Department of Quaternary Geology, Tornav\"{a}gen 13, SE-223 63 Lund}\\
{\em $^{\bullet}$Department of Nuclear Physics, Lund University}\\
{\em S\"olvegatan 14, SE-223 62 Lund, Sweden}\\
{\em $^{\circ}$Institute of Archaeology, Bucharest}\\}
\date{}
\begin{document}
\maketitle
\large
\thispagestyle{empty}
\begin{abstract}
In this study we have dated two human fossil remains found in Romania, 
by the method of radiocarbon using the technique of the accelerator mass spectrometry. 
The human fossil remains from  Woman's cave, Baia deFier, have been dated 
to the age 30150 $\pm$ 800 years BP, and the skull from the Cioclovina cave has been dated 
to the age 29000 $\pm$ 700 years BP. These are the most ancient dated till now human fossil remains 
from Romania, possibly belonging to the upper Paleolithic, the Aurignacian period.
\end{abstract}

\section{Introduction}
In this study we have dated two human fossil remains found in Romania, by the method of radiocarbon
dating using the accelerator mass spectrometry (AMS) technique, performed at the Pelletron 
accelerator in Lund University, Sweden.
These are the most ancient dated human fossil remains from Romania, attributed by some 
archaeologists to the upper Paleolithic, the Aurignacian period.
The first skull, and scapulum and tibia remains were found in 1952 in Baia de Fier, in the 
Woman's Cave, in Hateg, Gorj county in the province Oltenia, by Constantin Nicolaescu-Plopsor.
Another skull was found in Cioclovina cave, near commune Bosorod, Hunedoara county in Transylvania,
 found by a worker at the exploitation of phosphate deposits, in the year 1941. 
The skull arrived at Francisc Rainer, anthropologist, and Ioan Simionescu, geologist, 
who published a study of this skull.\cite{1}
The absence of stratigraphical  observation information made the cultural and chronological 
attribution of these skulls very difficult, and a number of archaeologists questioned  the 
Paleolithic character of these fossil remains. For this reason, the dating of the two skulls 
by a physical analysis is decisive.

\section{Experimental}
Samples of bone were taken from the scapulum and tibia remains from the Woman's cave, 
from Baia de Fier, and from the skull from the Cioclovina cave. 
The $^{14}$C content was determined in the two samples by using the AMS system at Lund University.  
Normally, sufficient collagen for AMS measurements can be extracted from bone fragments with 
masses of 1 g, or more provided that at least 5 to 10\% of the original collagen content is 
present. But for the presently studied bone remains, because of the small available quantity of 
 very old bone samples, the determination of the radiocarbon content in the bones was difficult. 
We have essentially applied the Longin method \cite{2} for the extraction of {\it{collagen}} 
from the bone structure.  We use the {\it{collagen}} to refer to collagen that has undergone a 
degree of diagenesis.  The next step is the transformation of the {\it{collagen}} into pure carbon 
 in an 
 experimental set-up for the preparation of samples for AMS technique \cite{3}. The pure carbon, 
 placed in a copper holder, is arranged in a wheel, together with two standards of oxalic acid and 
 one anthracite background sample. The wheel with the samples and standards is put into the ion 
 source of the accelerator.  The central part of the Lund AMS system is a Pelletron tandem 
 accelerator.  The accelerator is run at a terminal voltage of 2.4 MV during AMS experiments. 
The particle identification and measuring system consists of a silicon surface barrier detector 
 of diameter of 25 mm.
The computer system alternately analyses the data of the $^{13}$C current received from a 
 current integrator and the $^{14}$C counts arriving from the particle detector, to obtain, 
 finally, the ratio $^{14}$C/$^{13}$C for each sample. Each sample is measured 7 times. 
 The precision of the measurements for samples close to Modern \cite{4} is around 1 \% \cite{5}.

\section{Results and Archaeological Considerations}
Dating of the two sampes by the AMS-technique gave the following results:\\
\begin{center}
\begin{tabular}{lc}
\hline
\hline
Skull Woman's cave, Baia de Fier, Gorj, Oltenia &	30150 $\pm$ 800 years BP\\
\hline
Skull Cioclovina cave, Bosorod, Hunedoara	&       29000 $\pm$ 700 years BP\\
\hline
\hline
\end{tabular} 
\end{center}

\vspace*{0.5cm}

Fig. 1 shows the skull found alongside with fragments of scapulum, mandible and a dyaphisis of tibia in Baia de Fier, in the Woman's Cave, in Gorj county in the province Oltenia, by C. S. Nicolaescu-Plopsor in 1952.
C. S. Nicolaescu-Plopsor worked for the establishing of the geologic and archaeologic stratigraphy . He found two Paleolithic layers from the lithic pieces, one Musterian and one   
          Aurignacian.
In 1952 he found a skull, which he assumed to be a {\it{homo sapiens fossilis}}. The skull has been found in a gallery known as the "Musterian gallery". In the following years he and his collaborators discovered in a different area of the same cave fragments of scapulum, mandibula and tibia,  this archeological material being attributed to the Middle Paleolithic (Musterian). Even though the skull was found in the layer, the levels here are mixed and the stratigraphy is not clear. 
Most of the lithic materials discovered in the Woman's cave from Baia de Fier belong to the Musterian period. The few Aurignacian pieces from this cave have been discovered near the entrance of the cave, while the analyzed scapulum and tibia have been found inside the cave.
This circumstance renders difficult the association of the human remains with a specific historic cultural period.[6,7]
Fig. 2 shows the skull found in Cioclovina cave, near commune Bosorod, Hunedoara county in Transylvania, which was found by a worker at the exploitation of phosphate deposits, in the year 1941. The stratigraphical conditions in which the skull has been discovered were unclear from the moment of the discovery, and the association with the few lithic pieces found in the cave is not  established. The skull arrived at Francisc Rainer, anthropologist, and Ioan Simionescu, geologist,  who published a study of this skull.[8]
These authors advanced the hypothesis that the skull belongs to the man of the type {\it{Homo sapiens fossilis}}, with strong {\it Neanderthalian} characters. Unfortunately no further anthropological studies of this skull have been reported since then. Moreover, certain authors have questioned the Paleolithical age of this skull.
The skull is now in the custody of Theodor Neagu from the Faculty of Geology of the University of Bucharest.
In this condition, the dating of the two human fossil remains by physical analysis was crucial.

\section{Conclusions}
The analysis of the fossil remains from Baia de Fier and Cioclovina caves by radiocarbon using 
the AMS technique have demonstrated that the remains are very ancient and could be attributed to 
the period of upper Paleolithic period, the Aurignacian. On the basis of the dating of the fossil 
remains presented in this study, a future cultural identification might be possible, and in this 
way the fossil remains might be associated with other findings of the same type from the Central 
and Eastern Europe.

\noindent
Figure captions \\
\begin{tabular}{p{1.5cm}p{12cm}}

Fig. 1. & Skull found in the Women's cave, Baia de Fier Hateg, Gorj county in the province Oltenia \\

Fig. 2. & Skull found in the Cioclovina cave, near commune Bosorod, Hunedoara county in Transylvania \\
\end{tabular}
\end{document}